\documentclass[doublecol]{epl2}
\usepackage{amsfonts}

\usepackage{subfigure}
\usepackage{graphicx}
\usepackage{amsmath,amssymb}

\title{Solving the Cold-Start Problem in Recommender Systems with Social Tags}

\author{Zi-Ke ZHANG\inst{1} \and  Chuang LIU\inst{2,3} \and Yi-Cheng ZHANG\inst{1,4}\thanks{E-mail:\email{yi-cheng.zhang@unifr.ch}} \and Tao ZHOU\inst{4,5}}
\shortauthor{Zi-Ke ZHANG \etal}

\institute{
  \inst{1} Department of Physics, University of Fribourg, Chemin du Mus\'ee 3, 1700 Fribourg, Switzerland\\
  \inst{2} School of Business, East China University of Science and Technology, Shanghai
  200237, P. R. China \\
  \inst{3} Engineering Research Center of Process Systems Engineering (Ministry of Education), East China University
    of Science and Technology, Shanghai 200237, P. R. China \\
  \inst{4} Web Sciences Center, University of Electronic Science and Technology of China, Chengdu 610054, P. R. China \\
  \inst{5} Department of Modern Physics, University of Science and Technology of China, Hefei 230026, P. R. China
}


\abstract{In this Letter, based on the user-tag-object tripartite
graphs, we propose a recommendation algorithm that makes use of
social tags. Besides its low cost of computational time, the
experimental results on two real-world data sets, \emph{Del.icio.us}
and \emph{MovieLens}, show that it can enhance the algorithmic
accuracy and diversity. Especially, it provides more personalized
recommendation when the assigned tags belong to diverse topics.
The proposed algorithm is particularly effective for small-degree
objects, which reminds us of the well-known \emph{cold-start
problem} in recommender systems. Further empirical study shows that
the proposed algorithm can significantly solve this problem in
social tagging systems with heterogeneous object degree
distributions.}

\begin{document}

\maketitle

\begin{figure}
\begin{center}
\includegraphics[width=6.8cm]{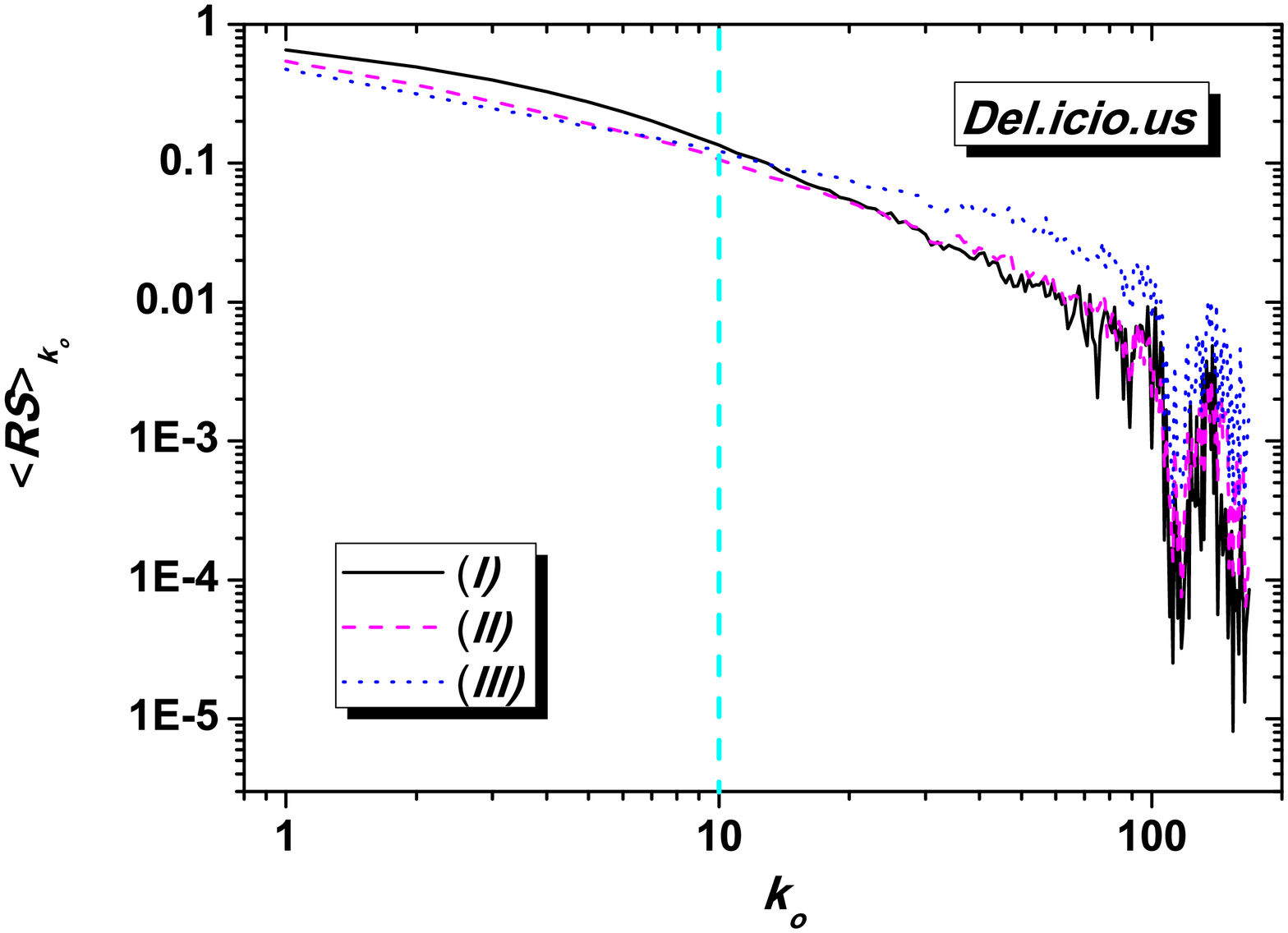}
\includegraphics[width=6.8cm]{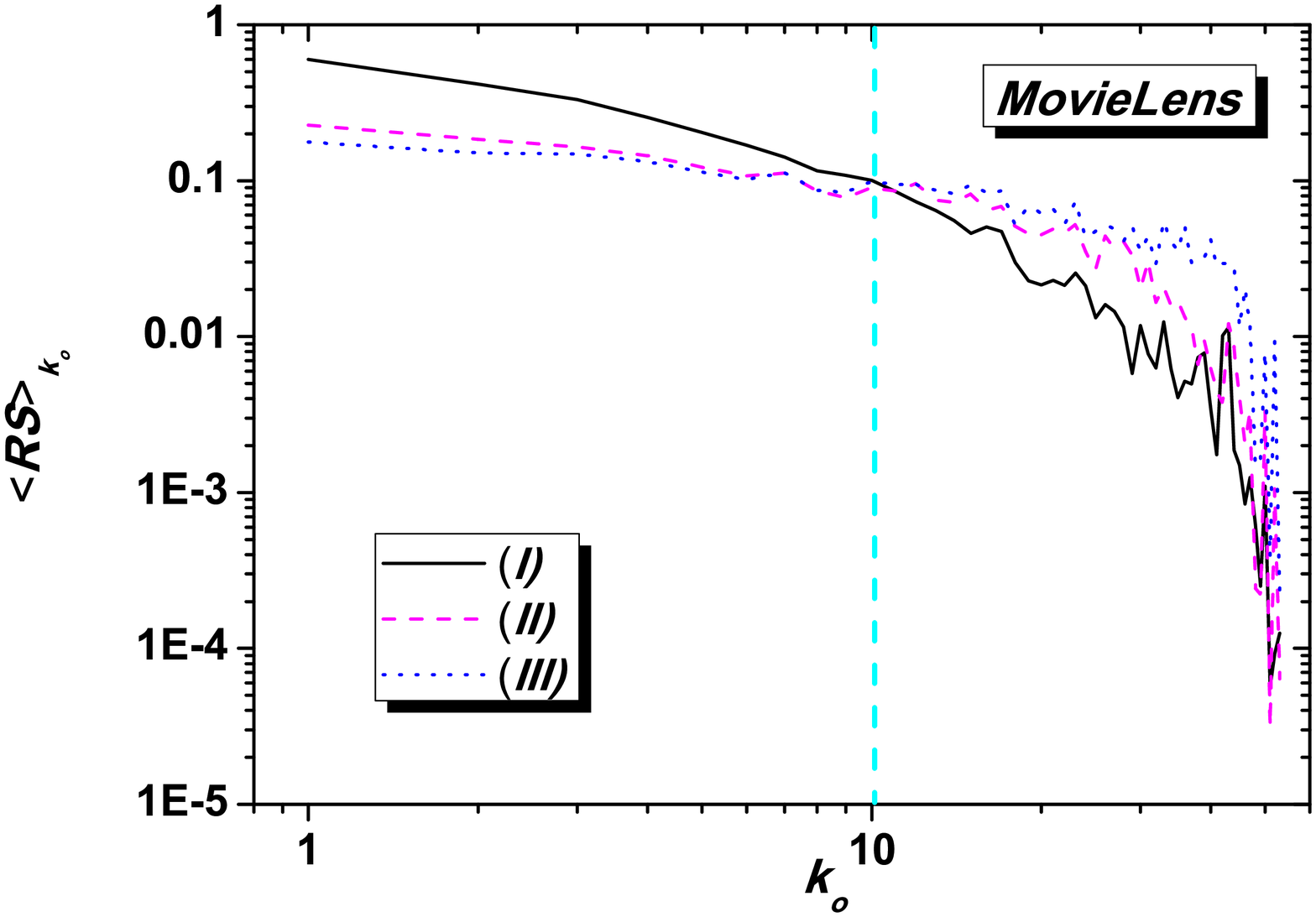}
\caption{(Color online) Object-degree-dependent ranking score for
the three algorithms in \emph{Del.icio.us} and \emph{MovieLens}.
Each data point is obtained by averaging over 50 realizations, each
of which corresponds to an independent division of training set and
testing set.}\label{coldstart}
\end{center}
\end{figure}

\begin{figure}
\begin{center}
\includegraphics[width=6.8cm]{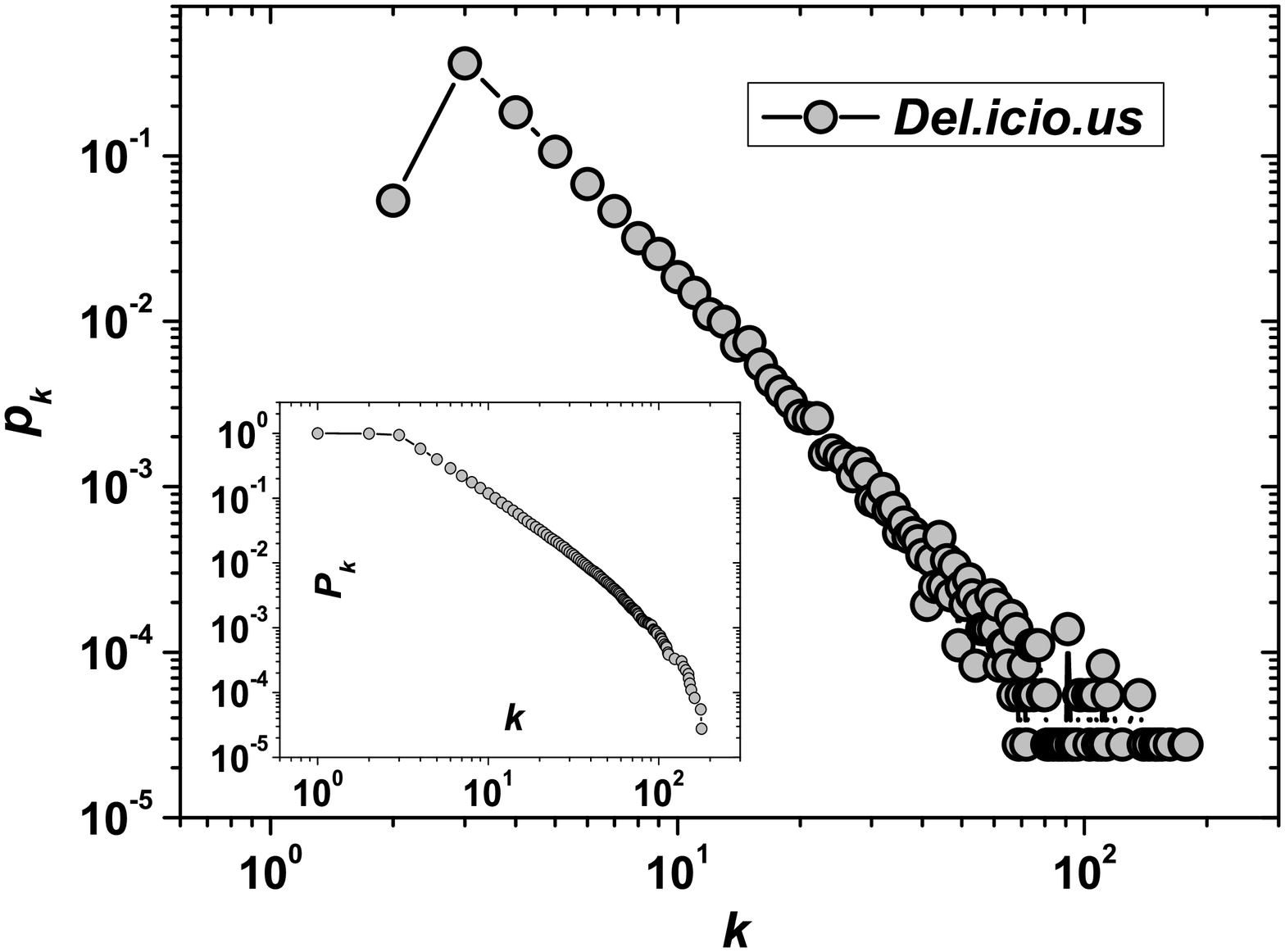}
\includegraphics[width=6.8cm]{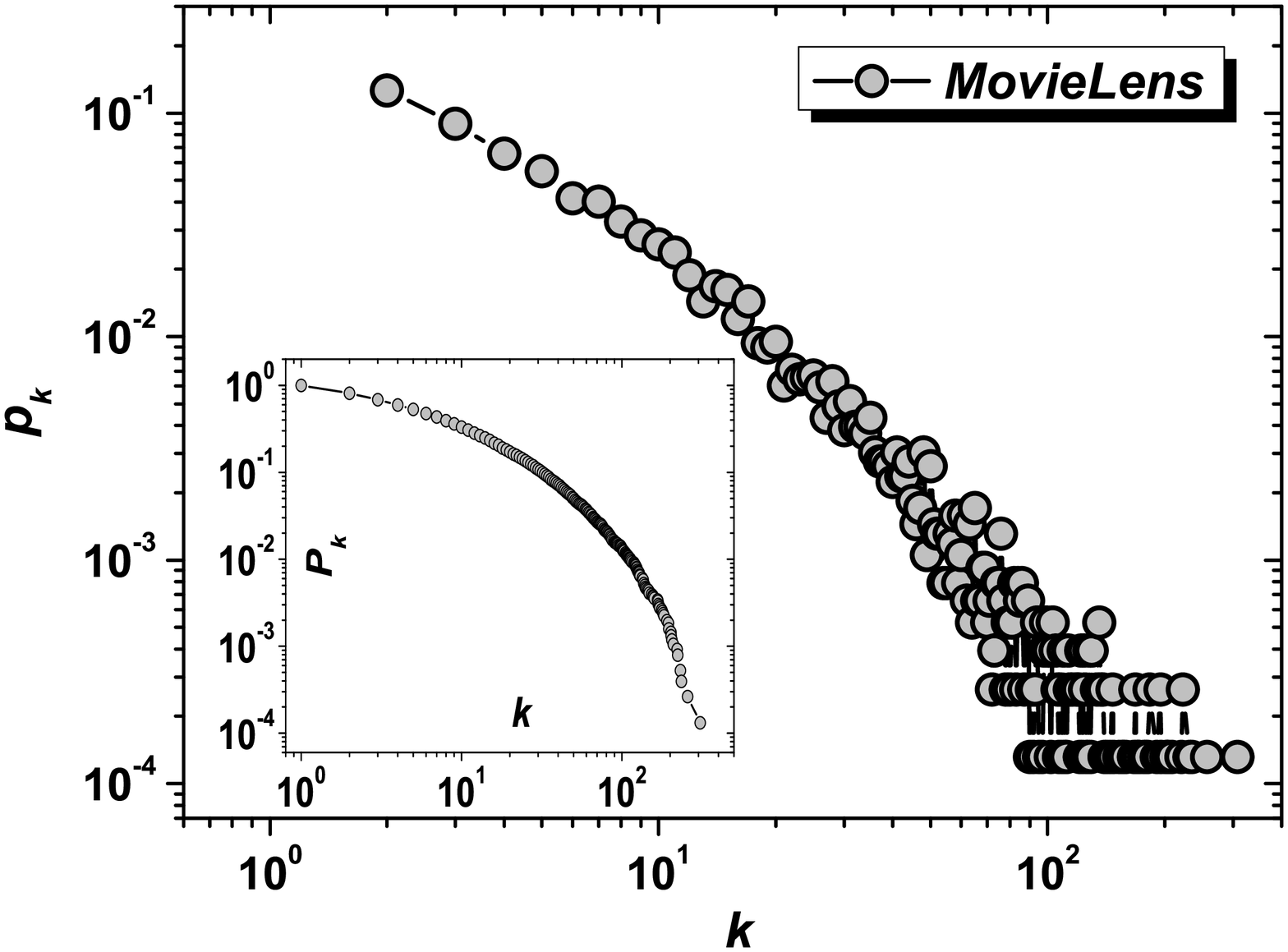}
\caption{ Object degree distributions of the two data sets. The
insets show the accumulative distributions}\label{degree}
\end{center}
\end{figure}

\section{Introduction}
Many complex systems can be well described by networks where nodes
represent individuals, and edges denote the relations among them
\cite{Albert2002, Dorogovtsev2002, Newman2003, Boccaletti2006,
Costa2007}. Recently, the personalized recommendation in complex
networks has attracted increasing attention from physicists
\cite{ZhangYC200701, ZhangYC200702, ZhouT2007, ZhouT200801,
ZhouT2010}. Personalized recommendation aims at finding objects
(e.g. books, webpages, music, etc.) that are most likely to be
collected by users. For example, classical information retrieval can
be viewed as recommending documents with given words
\cite{Salton1983}, and the process of link prediction can be
considered as a recommendation problem in unipartite networks
\cite{LvLY20091, LvLY2009, LiuWP2010}. The central problem of
personalized recommendation can be divided into two parts: one is
the estimation of similarity based on the historical records of user
activities \cite{Balabanovic1997, Sarwar2001}; the other is the
usage of accessorial information (e.g., object attributes) to
efficiently filter out irrelevant objects. For the formal task,
since computing and storing the similarities of all user pairs is
costly, we usually consider only the top-\emph{k} most similar users
\cite{Deshpande2004}. For the latter task, very accurate
descriptions of objects may be helpful in filtering irrelevant
objects, however, it is limited to the attribute vocabulary, and, on
the other hand, attributes describe global properties of objects
which are less helpful to generate personalized recommendations.

Recently, the advent of Web2.0 and its affiliated applications bring
a new form of paradigm, \emph{social tagging systems} (or called
\emph{collaborative tagging systems}), which introduces a novel
platform for users' participation. A social tagging system allows
users to freely assign tags to annotate their collections, requires
no specific skills for users to participate in, broadens the
semantic relations among users and objects, and thus has attracted
much attention from the scientific community. Golder \emph{et al.}
studied its usage patterns and classified seven kinds of tag
functions \cite{Golder2006}. Similar to the tagging functions, the
keywords and PACS numbers are analyzed to better characterize the
structure of co-authorship and citation networks \cite{Palla2008,
ZhangZK2008}. Furthermore, many efforts have been done to explain
the emergent properties of social tagging systems. Cattuto \emph{et
al.} \cite{Cattuto20071} proposed a memory-based Yule-Simon model to
describe the aging effects and occurrence frequencies of tags. Zhang
and Liu \cite{ZhangZK201002} proposed an evolutionary hypergraph
model, where users not only assign tags to objects but also retrieve
objects via tags.

Besides, social tagging systems have already found wide applications in
\emph{Recommender Systems}. By
considering the tag frequency as weight, Szomszor \emph{et al.}
\cite{Szomszor2007} proposed an improved movie recommendation
algorithm. Schenkel \emph{et al.} \cite{Schenkel2008} proposed an
incremental threshold algorithm taking into account both the social
ties among users and semantic relations of different tags, which
performs remarkably better than the algorithm without tag expansion.
Zhang \emph{et al.} \cite{ZhangZK201001} and Shang \emph{et al.}
\cite {ShangMS2010} proposed an object-based and user-based hybrid
tag algorithm, respectively, harnessing diffusion-based methods to
obtain better recommendations. Shang and Zhang \cite{ShangMS2009}
considered the tag usage frequency as edge weight in a user-object bipartite
network and improved the accuracy of recommendation.

In this Letter, we propose a diffusion-based recommendation
algorithm which considers social tags as a bridge connecting users
and objects. That is to say, users can efficiently find the target
objects via tags. In particular, we consider the usage frequencies
of tags as users' personal preference, while the semantic relations
between tags and objects as global information. Experimental results
show that the present algorithm can significantly improve the
recommendation accuracy. Further empirical study shows that the
proposed algorithm is especially effective for the objects collected
by few users, which reminds us of the well-known \emph{cold-start}
problem \cite{Schein2001, Schein2002}. Since there is little
information available for new objects, social tags can effectively
build up relations between existing objects and the new ones.
Therefore, the incorporating of tags can remarkably help users find
the new (or less popular) yet interesting objects, and thus enhance
the overall accuracy. In addition, we employ entropy-based and
Hamming-distance-based methods to measure the \emph{inner-} and
\emph{inter-} diversity of tag usage patterns, respectively.
Experimental results show that there are different tag usage
patterns in the two datasets: users assign more
diverse tags in \emph{Del.icio.us} than \emph{MovieLens}, and it
might shed lights on understanding why the proposed algorithm can
enhance the recommendation diversity in \emph{Del.icio.us} largely
than \emph{MovieLens}.


\section{Data}
\begin{figure}[ht]
\begin{center}
\includegraphics[width=6.8cm]{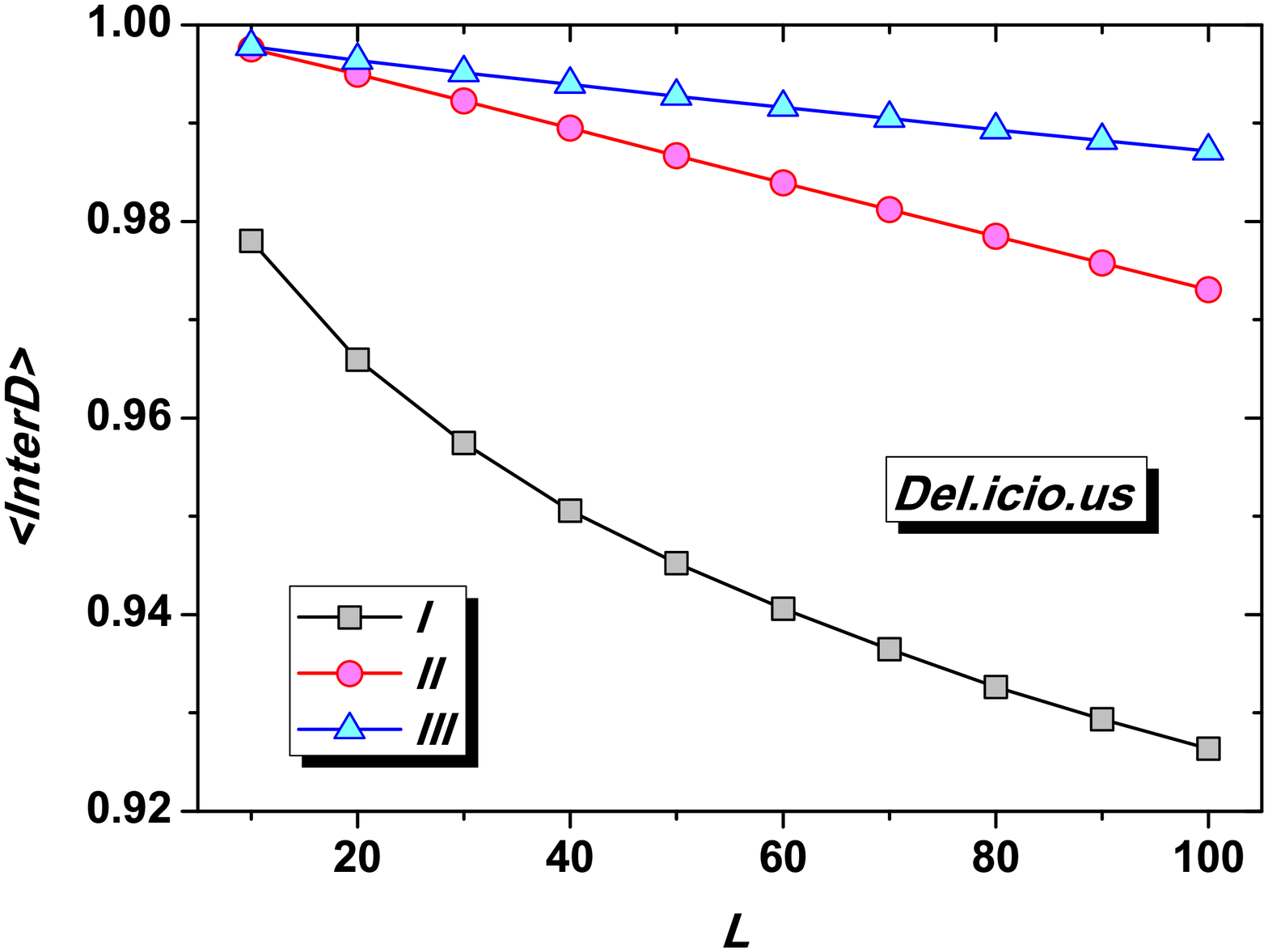}
\includegraphics[width=6.8cm]{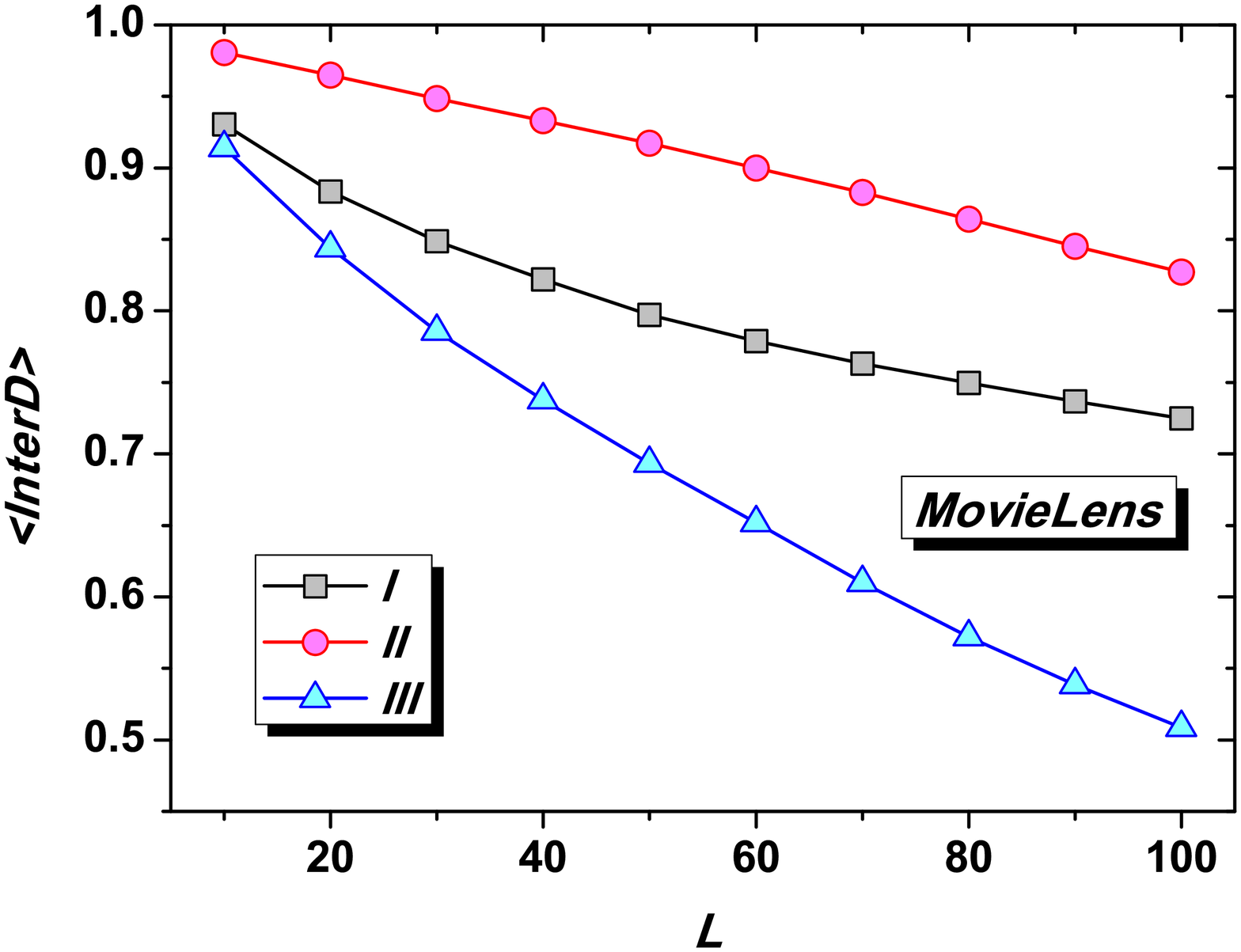}
\caption{(Color online) $\langle InterD\rangle$ as a function of the
length of recommendation list for the three algorithms in
\emph{Del.icio.us} and \emph{MovieLens}. }\label{interdiverstiy}
\end{center}
\end{figure}

\begin{figure}
\begin{center}
\includegraphics[width=6.8cm]{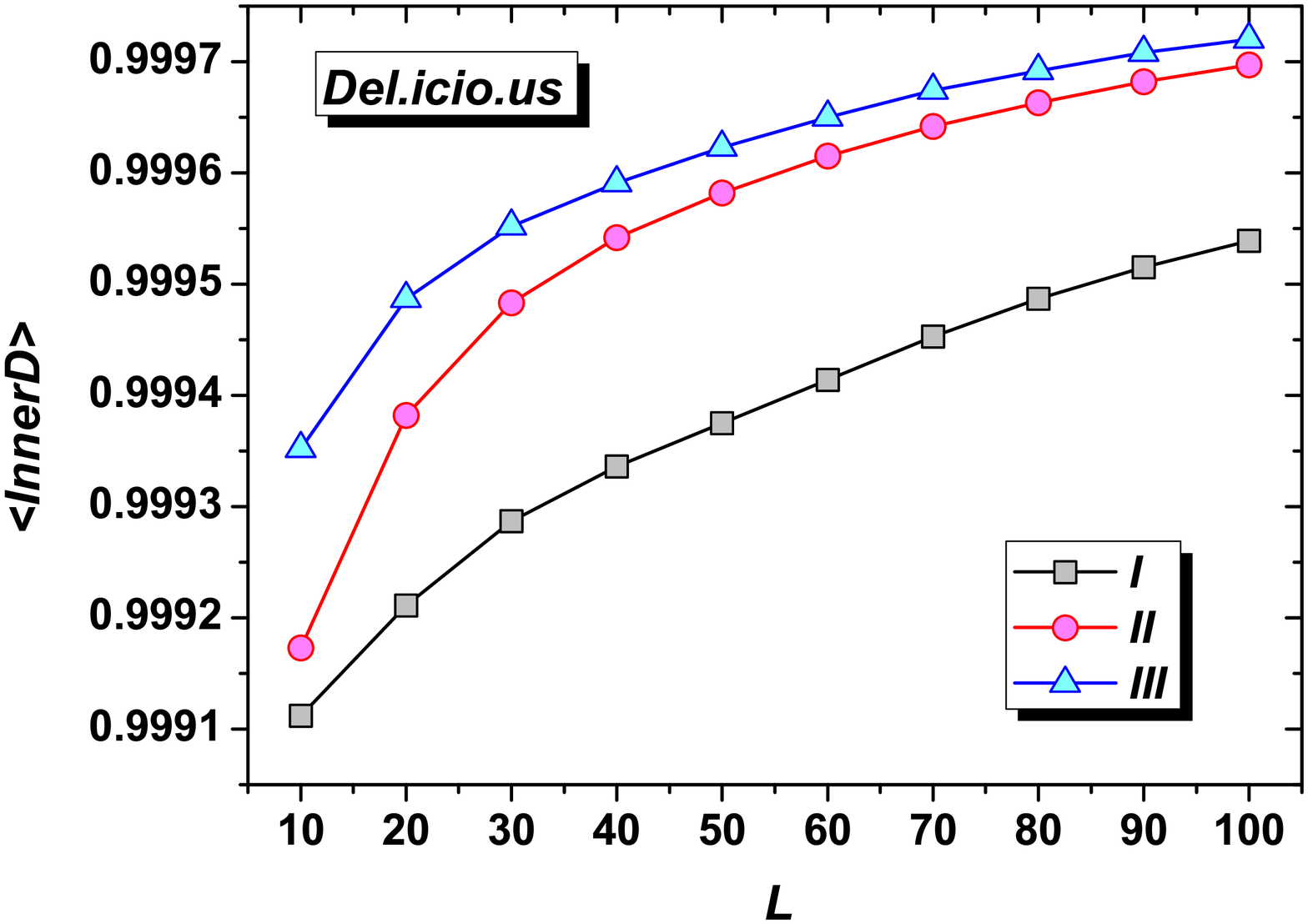}
\includegraphics[width=6.8cm]{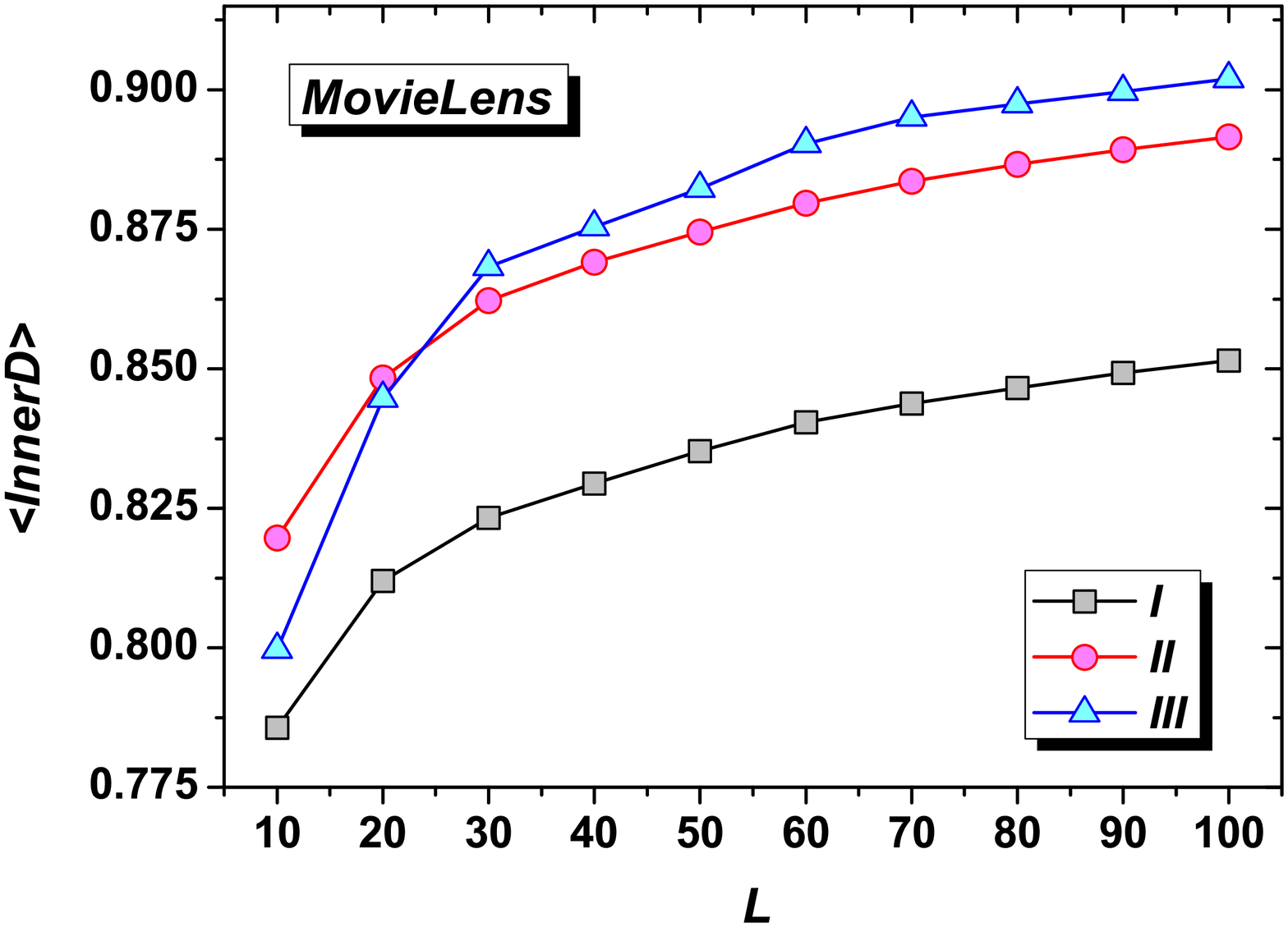}
\caption{(Color online) $\langle InnerD\rangle$ as a function of the
length of recommendation list for the three algorithms in
\emph{Del.icio.us} and \emph{MovieLens}. }\label{innerdiverstiy}
\end{center}
\end{figure}
The empirical data used in this paper include: (i)
\emph{Del.icio.us} --one of the most popular social bookmarking web
sites, which allows users not only to store and organize personal
bookmarks (URLs), but also to look into other users' collections and
find what they might be interested in by simply keeping track of the
baskets with social tags; (ii) \emph{MovieLens} --a movie rating
system, where each user votes movies in five discrete ratings 1-5. A
tagging function is added in from January 2006. In both data sets,
we remove the isolated nodes and guarantee that each user has
collected at least one object, each object has been collected by at
least two users, assigned by at least two tags, and each tag is used
by at least by two users, and each tag is used at least twice by
every adjacent user. Table 1 summarizes the basic statistics of the
purified data sets.

\begin{table}
\centering \caption{Basic statistics of the two data sets. $n$, $m$,
$r$ are the total numbers of users, objects and tags, respectively.
$\langle k\rangle$, $\langle k'\rangle$ and $\langle k''\rangle$
denote the average number of objects collected by a user, tags
assigned by an object and tags adopted by a user respectively.
\emph{Del.} and \emph{Mov.} represent the data sets
\emph{Del.icio.us} and \emph{MovieLens}, respectively.}
\begin{tabular}{ccccccc}  \hline  Data & $n$ & $m$ & $r$ & $\langle k\rangle$ & $\langle
k'\rangle$ & $\langle k''\rangle$
\\ \hline
\emph{Del.} &4902&36224&  10584 &  43.85 &  38.82 & 286.86 \\
\emph{Mov.} &648&1590  &1382 &  15.04 &  19.89 &  22.89 \\
\hline
\end{tabular}
\end{table}

Every data set is consisted of many entries, and each follows the
form $\mathbb{F}$=\{user, object, tag$_1$, tag$_2$, $\cdots$,
tag$_t$\}, where $t$ is the number of tags assigned to this object
by this user. Then each data set is randomly divided into two parts:
the training set, is treated as known information, while the testing
set is used for testing. In this Letter, the training set always
contains 90\% of entries and the remaining 10\% of entries
constitute the testing set.


\begin{figure}
\begin{center}
\includegraphics[width=6.8cm]{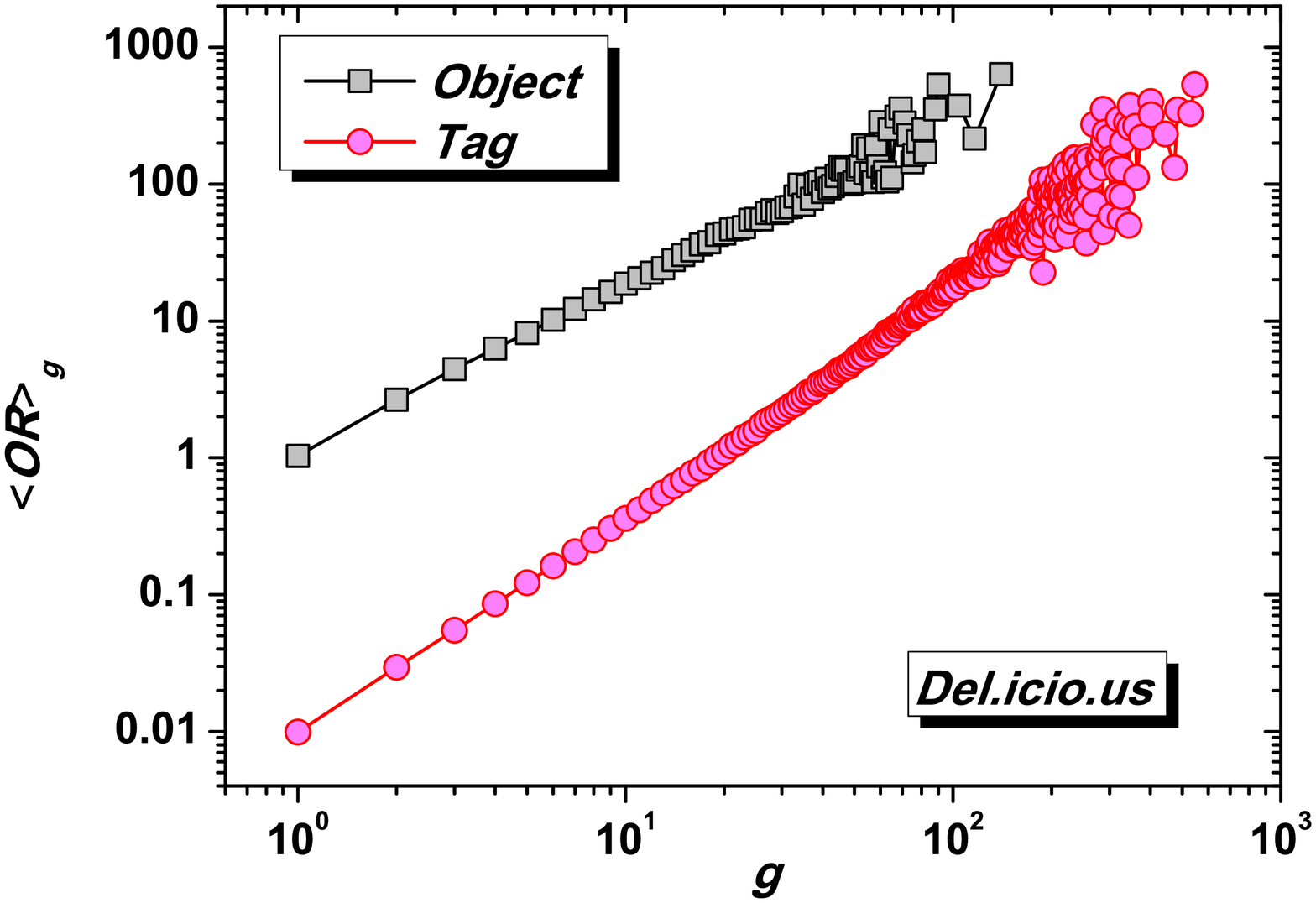}
\includegraphics[width=6.8cm]{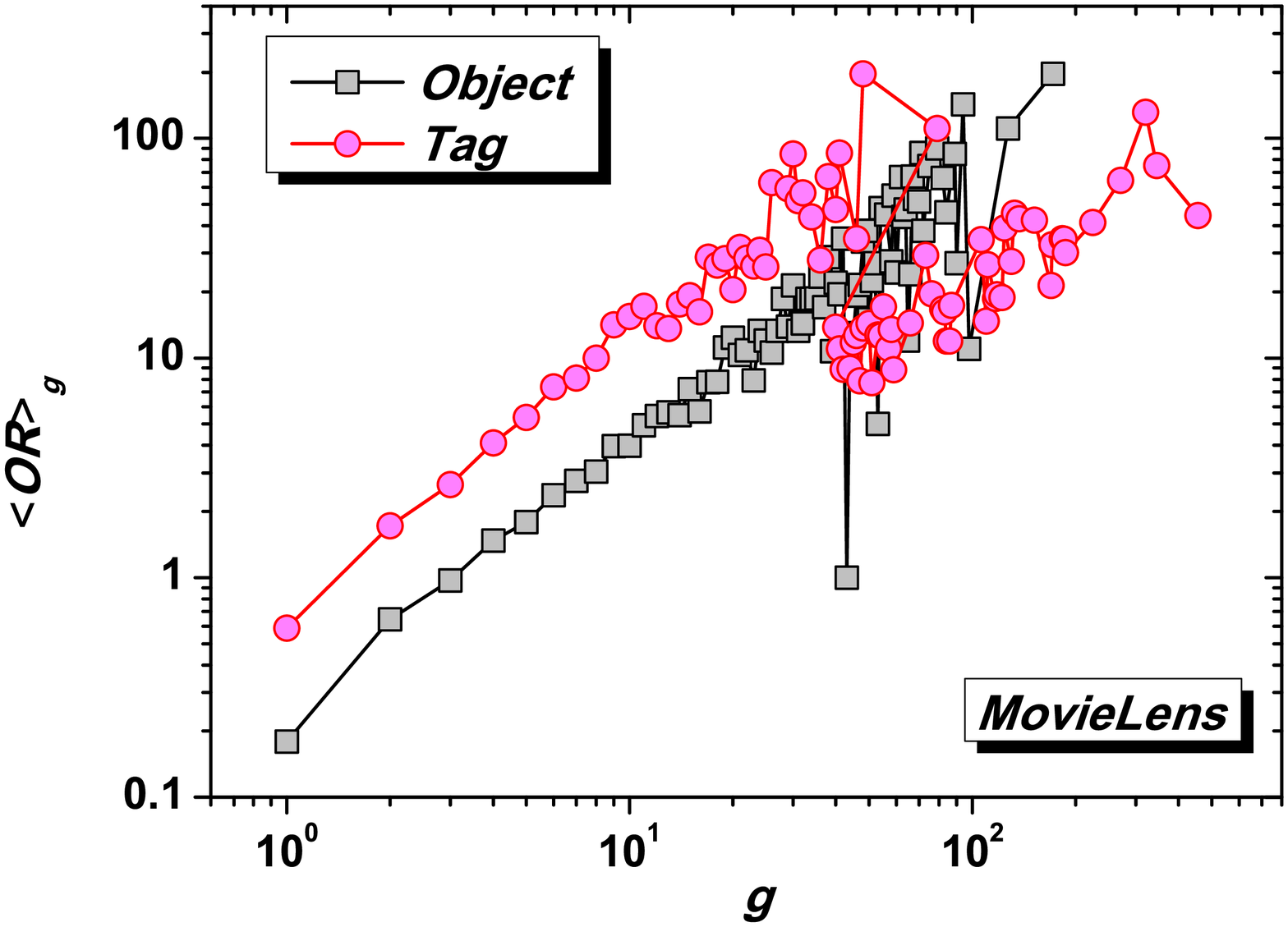}
\caption{(Color online)  $\langle OR\rangle$ as a function of $g$
for the two data sets. The black squares represent $\langle
OR\rangle$ for objects and the red circles are $\langle OR\rangle$
of tags, respectively. }\label{hamming}
\end{center}
\end{figure}

\begin{figure}[ht]
\begin{center}
\includegraphics[width=6.8cm]{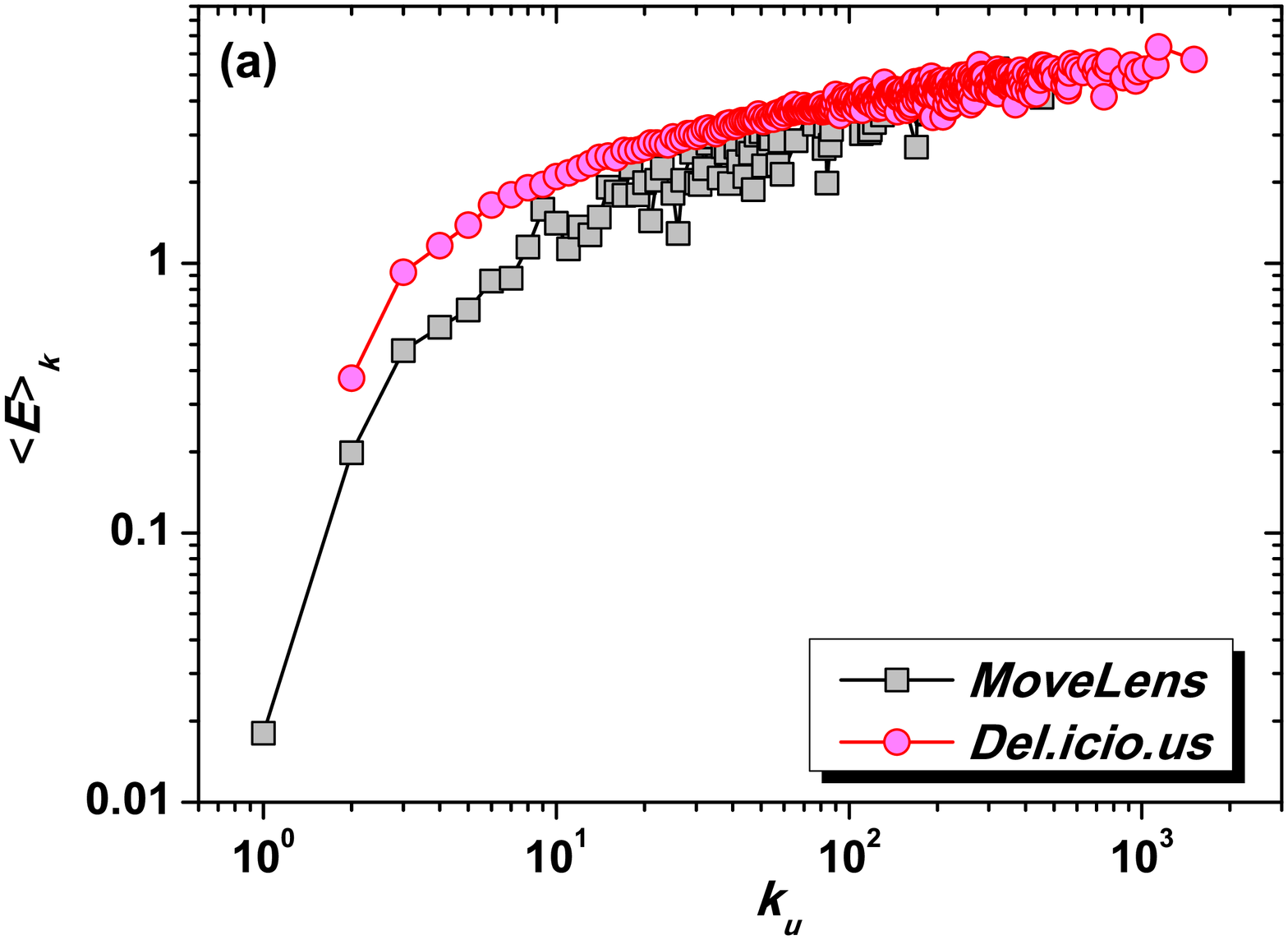}
\includegraphics[width=6.8cm]{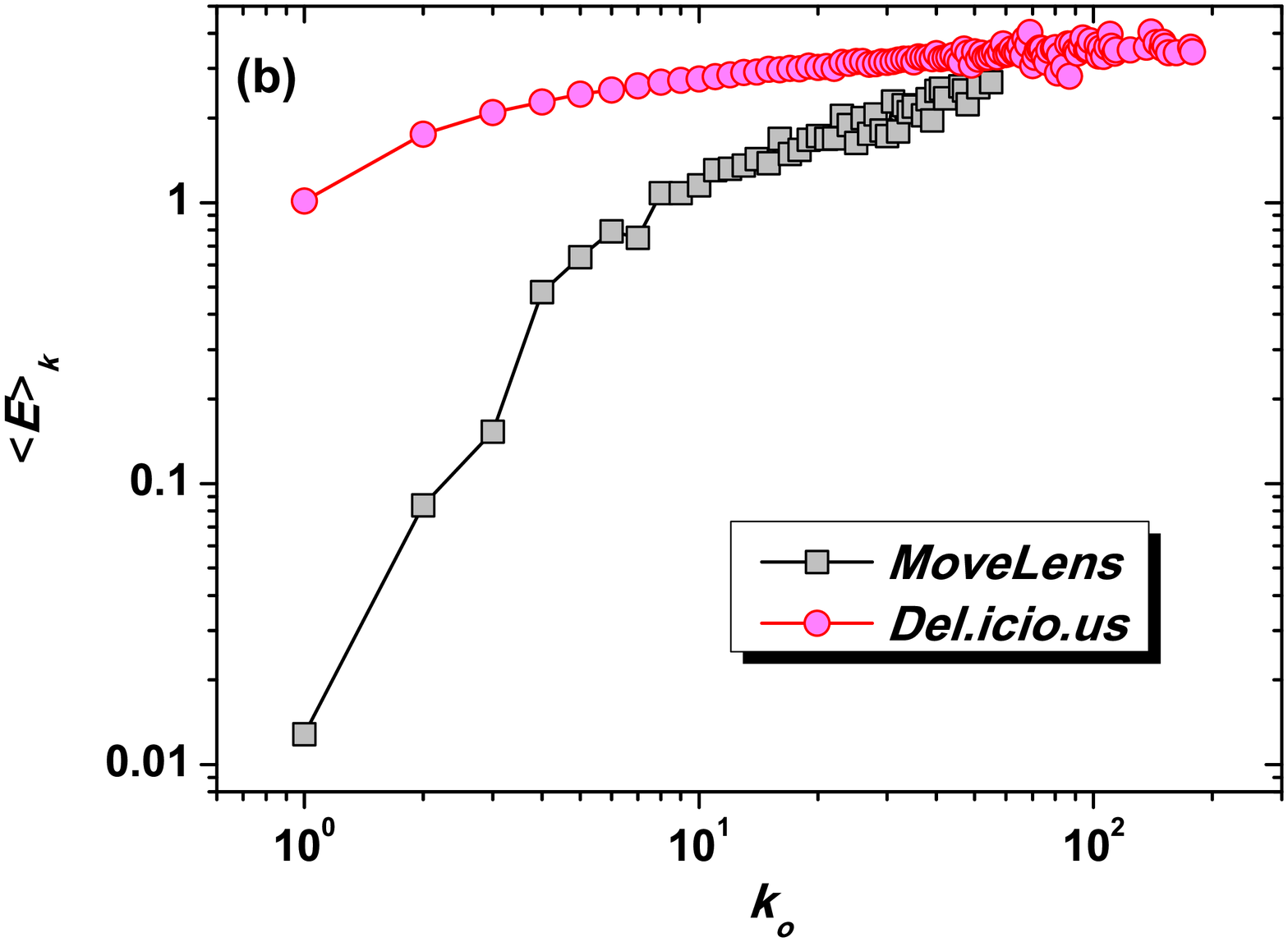}
\caption{(Color online) (a) $\langle E\rangle$ as a function of user
degree; (b) $\langle E\rangle$ as a function of object degree,
respectively. }\label{entropy}
\end{center}
\end{figure}

\section{Algorithms}
A recommender system considered in this Letter consists of three
sets, respectively of users \emph{U} =
\{$U_{1}$,$U_{2}$,$\cdots$,$U_{n}$\}, objects \emph{O} =
\{$O_{1}$,$O_{2}$,$\cdots$,$O_{m}$\}, and tags \emph{T} =
\{$T_{1}$,$T_{2}$,$\cdots$,$T_{r}$\}. The tripartite graph
representation can be described by three matrices, $A$, $A'$ and
$A''$ for user-object, object-tag and user-tag relations. If $U_i$
has collected $O_j$, we set $a_{ij}$ = 1, otherwise $a_{ij}$ = 0.
Analogously, we set $a'_{jk}$ = 1 if $O_j$ has been assigned by the
tag $T_k$, and $a'_{jk}$ = 0 otherwise. Furthermore, the users'
preferences on tags can be represented by a weighted matrix $A''$,
where $a''_{ik}$ is the number of times that $U_i$ has adopted
$T_k$.

Subsequently, we introduce the proposed algorithm, as well as two
baseline ones: (I) user-object diffusion \cite{ZhouT2007}; (II)
user-object-tag diffusion \cite{ZhangZK201001}; (III)
user-tag-object diffusion. Given a target user $U_i$, the above
three algorithms will generate final score of each object, $f_{j}$,
that are pushed into recommendation resource for him/her, are
described as following:

(I) Supposing that a kind of resource is initially located on
objects. Each object averagely distributes its resource to all
neighboring users, and then each user redistributes the received
resource to all his/her collected objects. The final resource vector
for the target user $U_i$, $\vec{f}$, after the two-step diffusion
is:
\begin{equation}
 f_j=\sum_{l=1}^n\sum_{s=1}^m\frac{a_{lj}a_{ls}a_{is}}{k(U_l)k(O_s)},
 \ j=1,2,\cdots,m,
\end{equation}
where $k(U_l)=\sum_{j=1}^ma_{lj}$ is the number of collected objects
for user $U_l$, and $k(O_s)=\sum_{i=1}^na_{is}$ is the number of
neighboring users for object $O_s$.

(II) The initial resources are set as same as I, but each object
equally distributes its resource to all neighboring tags, and then
each tag redistributes the received resource to all its neighboring
objects. Thus, the final resource vector, $\vec{f'}$, is:

\begin{equation}
f'_j=\sum_{l=1}^r\sum_{s=1}^m\frac{a'_{jl}a'_{ls}a{_{is}}}{k'(T_l)k'(O_s)}
 \ j=1,2,\cdots,m,
\end{equation}
where $k'(T_l)=\sum_{j=1}^ma'_{jl}$ is the number of neighboring
objects for tag $T_l$, $k'(O_s)=\sum_{l=1}^ra'_{sl}$ is the number
of neighboring tags for object $O_s$.

(III) Different from I and II, here, the initial resources are
located on tags according to their frequencies used by the target
user $U_i$. Then each tag distributes the initial resource directly
to all its neighboring objects. Thus, the final resource vector,
$\vec{f''}$, reads:

\begin{equation}
 f''_j=\sum_{l=1}^r\frac{a'_{jl}a''_{il}}{k'(T_l)}
\end{equation}

After we obtain the final score of objects, all the objects that
$U_i$ has not collected are ranked in a descending order, and the
top $L$ objects will be recommended to $U_i$.

Comparing with algorithms I and II, the advantages of algorithm III
are threefold. Firstly, since social tags highly reflect users'
personal preferences, algorithm III is promisingly expected to
generate more personalized recommendation. Secondly, the one-step
diffusion can clearly save computational time especially for
large-scale data. Thirdly, algorithm III reveals the essential role
of tags: building a bridge between users and objects, helping users
retrieve and organize collections without the limit of hierarchial
structure and vocabulary of words.

\begin{table}
\centering \caption{Algorithmic accuracy for \emph{Del.icio.us}.
$\langle RS\rangle_{k_{o}\leq10}$ is the average ranking score over
objects with degree equal or less than 10, and $\langle
RS\rangle_{k_{o}>10}$ is the average ranking scores over objects
with degree greater than 10. Each value is obtained by averaging
over 50 realizations, each of which corresponds to an independent
division of training set and testing set.}
\begin{tabular}{cccc}  \hline Algorithms & $\langle RS\rangle$ & $\langle RS\rangle_{k_{o}\leq10}$ & $\langle RS\rangle_{k_{o}>10}$  \\
\\ \hline
 I  & 0.276 & 0.369 & 0.054 \\
 II & 0.209 & 0.275 & 0.049 \\
 III & 0.196 & 0.249 & 0.068 \\
\hline
\end{tabular}
\end{table}

\begin{table}
\centering \caption{Algorithmic accuracy for \emph{MovieLens}. }
\begin{tabular}{cccc}  \hline Algorithms & $\langle RS\rangle$ & $\langle RS\rangle_{k_{o}\leq10}$ & $\langle RS\rangle_{k_{o}>10}$  \\
\\ \hline
 I & 0.207 & 0.307 & 0.039 \\
 II & 0.130 & 0.168  & 0.055\\
 III & 0.123 & 0.146 & 0.070 \\
\hline
\end{tabular}
\end{table}

\section{Metrics}

To give solid and comprehensive evaluation of the proposed
algorithm, we employ three different metrics that characterizing the
accuracy and diversity of recommendations.

\begin{enumerate}
\item \emph{Ranking Score} ($RS$) \cite{ZhouT2007}.---
In the present case, for each entry in the testing set (i.e. a user-object pair), $RS$
is defined as the rank of the object, divided by the number of all
uncollected objects for the corresponding user.
Apparently, the less the $RS$, the higher accuracy the algorithm is.
The average ranking score $\langle RS \rangle$ is given by averaging
over all entries in the testing set.

\item \emph{Inter Diversity} ($InterD$) \cite{ZhouT200802,
ZhouT2007}.--- $InterD$ measures the differences of different users'
recommendation lists, thus can be understood as the inter-user
diversity. Denote $O^i_R$ the set of recommended objects for user
$U_i$, then
\begin{equation}
    \emph{InterD} =
    \frac{2}{n(n-1)}\sum_{i\neq j}\left(1-\frac{|O^i_R\cap O^j_R|}{L}\right),
\end{equation}
where $L=|O^i_R|$  is the length of recommendation list. In average,
greater or less $InterD$ mean respectively greater or less
personalization of users' recommendation lists.

\item \emph{Inner Diversity} ($InnerD$) \cite{ZhouT200802}.---
$InnerD$ measures the differences of objects within a user's
recommendation list, thus can be considered as the inner-user
diversity. It reads,
\begin{equation}
    \emph{InnerD} =
    1-\frac{2}{nL(L-1)}\sum^n_{i=1}\sum_{j\neq l,j,l\in O^i_R}S_{jl},
\end{equation}
where
$S_{jl}=\frac{|\Gamma_{O_j}\cap\Gamma_{O_l}|}{\sqrt{|\Gamma_{O_j}|\times
|\Gamma_{O_l}|}}$ is the cosine similarity between objects $O_j$ and
$O_l$, where $\Gamma_{O_j}$ denotes the set of users having
collected object $O_j$. In average, greater or less $InnerD$
suggests respectively greater or less topic diversification of
users' recommendation lists.
\end{enumerate}

\section{Results}
To make clear the role of social tags, a microscopic picture of
algorithmic accuracy is very helpful. Especially, since social tags
are used to describe the objects, we would like to see the
dependence of accuracy on object degree, namely the number of users
collecting it. Given an object degree $k_o$, the degree-dependent
average ranking score, denoted by $\langle RS\rangle_{k_{o}}$, is
defined as the mean positions averaged over all the entries in the
testing set with object degree equal to $k_o$.

In Table 2 and Table 3, we give the overall $\langle RS\rangle$ of
the three algorithms for the observed data sets. It indicates that
the $\langle RS\rangle$ is significantly enhanced by the present
algorithm. Fig. \ref{coldstart} reports the correlation between
accuracy and object degree. The ranking score decays with the
increasing $k_o$ for all the three algorithms. In addition, the
three curves intersect around $k_o$=10, which is a relatively small
value considering the heterogeneous object-degree distribution shown
in Fig. \ref{degree}. From Fig. \ref{coldstart}, it is seen that the
algorithmic accuracy of algorithm III is better than that of
algorithms I or II for $k_o\leq$10, but worse when $k_o>$10 (see also
Table 2 and Table 3), which reminds us of the well-known cold-start
problem in recommender systems: how to recommend the unpopular
and/or new objects to users? It is very difficult for a user to be
aware of these \emph{cold objects} by random surfing since they are
not hot items, and for a recommender system to recommend them to
right places since there are usually insufficient information about
them. In fact, there are 90.04\% and 69.35\% objects with
$k_o\leq$10 in \emph{Del.icio.us} and \emph{MovieLens},
respectively. Therefore, a successful recommender system has to make
reasonable recommendations of cold objects. Comparing with the
algorithms I and II, the present one can effectively help users find
those cold objects via social tags.

Fig. \ref{interdiverstiy} and Fig. \ref{innerdiverstiy} show the
experimental results of $\langle InterD\rangle$ and $\langle
InnerD\rangle$, respectively. In Fig. \ref{interdiverstiy}, $\langle
InterD\rangle$ is enhanced only for \emph{Del.icio.us}. The reason
for small $\langle InterD\rangle$ of algorithm III in
\emph{MovieLens} is that there are only movies in that data set, and
thus a comparatively small number of tags are used with huge
overlapping. The overlapping ratio, $OR$, of tags for users to
assign to the same objects, is defined as:

\begin{equation}
    OR_{g} = \frac{1}{N_g}\sum_{i\neq j, G(i,j)=g}OR(i,j),
\end{equation}
where $N_g$ is the number of user pairs $(i,j)$ such that $i\neq j$,
and $G(i,j)=g$ denotes the number of common objects collected by
users $i$ and $j$. $OR(i,j)$ is defined as the total number of tag
agreements on the same objects for user pair $(i,j)$. Similar
definition can also be used to quantity the overlapping ratio of
objects collected by users with the same tags. Clearly, larger $OR$
indicates smaller diversity, and vice versa. Fig. \ref{hamming} shows
the correlation between $\langle OR\rangle_g$ and $g$. One can see
that $\langle OR\rangle_g$ of tags is smaller than that of objects
in \emph{Del.icio.us}, while it is not the case for
\emph{MovieLens}. In a word, social tags can help generate more
diverse recommendation only if the tags are themselves used in a
diverse way.

Fig. \ref{innerdiverstiy} shows that $\langle InnerD\rangle$ is
generally improved by our proposed algorithm, indicating that it can
help users broaden their horizons. Except for \emph{MovieLens} with
very small $L$. It is again resulted from the narrow choice of tags
in \emph{MovieLens}. Recently, the \emph{Shannon entropy} is widely
used to quantify network diversity in social sharing networks
\cite{Lambiottea2006} and social economics \cite{Eagle2010}. In the
Letter, we also employ it to measure individual usage pattern of
tags:

\begin{equation}
    E\left(U_i\right) = -\sum_t p_{i;t}\textmd{ln}(p_{i;t}),
\end{equation}
where $p_{i;t}$ is the probability for tag $t$ used by user $U_i$.
Then the dependence of entropy on user degree, $E_k$, is given by
averaging all the $E\left(U_i\right)$ with $k\left(U_i\right)=k$.
Similar definition can be used to quantify the dependence of entropy
for objects. Clearly, Larger $E_k$ means that the users are more
willing to use diverse topics of tags, or the objects are more
likely to be assigned to more diverse tags, and vice versa. Fig.
\ref{entropy} shows that $E$ of \emph{Del.icio.us} are greater than
that of \emph{MovieLens} for both users and objects, indicating that
\emph{Del.icio.us} is a more diverse system than \emph{MovieLens},
and further giving a reasonable explanation why algorithm III can
obtain better $InnerD$ in \emph{Del.icio.us} than \emph{MovieLens}.

\section{Conclusions and Discussion}
In this Letter, we proposed a recommendation algorithm making use of
social tags. This algorithm, considers the frequencies of tags as
user preferences on different topics and tag-object links as
semantical relations between them. Experimental results demonstrated
that the proposed algorithm outperforms the two baseline algorithms
in both accuracy and diversity.
The present algorithm
outperforms others especially for the objects with small degrees
($k_o\leq10$), which constitute the majority of objects. Therefore,
the incorporating of social tags could be, to some extent, helpful
in solving the cold-start problem of recommender systems. 

Recently, besides the accuracy, the significance of diversity has
attracted more and more attention in information filtering
\cite{ZhouT2010}. Experimental results in this Letter demonstrated
that a wide-range adoption of social tags can enhance the diversity
of recommendation. Therefore, we strongly encourage recommender
systems to add tagging functions and users to organize their
collections by using tags. However, despite the significant role of
tags, the polysemy and synonymy problems \cite{Golder2006} might
result in coarse and inaccurate performance, the tag clustering
technique \cite{Shepitsen2008} is hopefully to provide a promising
way to generate multi-scale recommendations and eventually obtain
the best performance.

\acknowledgments This work is partially supported by the Swiss
National Science Foundation (Project 200020-121848). Z.-K.Z. and
T.Z. acknowledge the National Natural Science Foundation of China
under the grant no. 60973069.


\begin{thebibliography}{0}

\bibitem{Albert2002}
  \Name{Albert R. \and Barab\'{a}si A.-L.}
  \REVIEW{Rev. Mod. Phys.}{74}{2002}{47}.

\bibitem{Dorogovtsev2002}
  \Name{Dorogovtsev S. N. \and Mendes J. F. F.}
  \REVIEW{Adv. Phys.}{51}{2002}{1079}.

\bibitem{Newman2003}
  \Name{Newman M. E. J.}
  \REVIEW{SIAM Rev.}{45}{2003}{167}.

\bibitem{Boccaletti2006}
  \Name{Boccaletti S., Latora  V., Moreno Y. , Chavez M. \and Huang D.-U.}
  \REVIEW{Phys. Rep.}{424}{2006}{175}.

\bibitem{Costa2007}
  \Name{Costa L. da F., Rodrigues F. A., Traviesor G. \and Boas P. R. U.}
  \REVIEW{Adv. Phys.}{56}{2007}{167}.



\bibitem{ZhangYC200701}
  \Name{Zhang Y.-C., Blattner M. \and Yu Y.-K.}
  \REVIEW{Phys. Rev. Lett.}{99}{2007}{154301}.

\bibitem{ZhangYC200702}
  \Name{Zhang Y.-C., Medo M., Ren J., Zhou T., Li T. \and Yang F.}
  \REVIEW{EPL}{80}{2007}{68003}.

\bibitem{ZhouT2007}
  \Name{Zhou T., Ren J., Medo M. \and Zhang Y.-C.}
  \REVIEW{Phys. Rev. E}{76}{2007}{046115}.

\bibitem{ZhouT200801}
  \Name{Zhou T., Jiang L.-L., Su R,-Q. \and Zhang Y.-C.}
  \REVIEW{EPL}{81}{2007}{58004}.

\bibitem{ZhouT2010}
  \Name{Zhou T., Kuscsik Z., Liu J.-G., Medo M., Wakeling J. R. \and Zhang Y.-C.}
  \REVIEW{Proc. Natl. Acad. Sci. U.S.A.}{107}{2010}{4511}.

\bibitem{Salton1983}
  \Name{Salton G. \and McGill M. J.}
  \emph{Introduction to Model Information Retrieva} (MuGraw-Hill, Auckland, 1983).

\bibitem{LvLY20091}
  \Name{Zhou T., L\"{u} L. \and Zhang Y.-C.}
  \REVIEW{Eur. Phys. J. B}{71}{2009}{623-630}

\bibitem{LvLY2009}
  \Name{L\"{u} L. \and Zhou T.}
  \REVIEW{EPL}{89}{2010}{18001}.

\bibitem{LiuWP2010}
  \Name{Liu W. \and L\"{u} L.}
  \REVIEW{EPL}{89}{2010}{58007}.

\bibitem{Balabanovic1997}
  \Name{Balabanovi\'{c} M. \and Shoham Y.}
  \REVIEW{Commun. ACM}{40}{1997}{72}.

\bibitem{Sarwar2001}
  \Name{Sarwar B., Karypis G., Konstan J.\and Riedl J.}
  \emph{Proc. the 10th Intl. Conf. WWW} (ACM Press, New York) 2001, pp. 295-305.

\bibitem{Deshpande2004}
  \Name{Deshpande M. \and Karypis G.}
  \REVIEW{ACM Trans. Inf. Syst.}{22}{2004}{143}.

\bibitem{Golder2006}
  \Name{Golder S. A. \and Huberman B. A.}
  \REVIEW{J. Info. Sci.}{32}{2006}{198}.

\bibitem{Palla2008}
  \Name{Palla G., Farkas I. J., Pollner P., Der\'{e}yi I. \and Vicsek T.}
  \REVIEW{New J. Phys.}{10}{2008}{123026}.

\bibitem{ZhangZK2008}
  \Name{Zhang Z.-K., L\"{u} L., Liu J.-G. \and Zhou T.}
  \REVIEW{Eur. Phys. J. B}{66}{2008}{557}.

\bibitem{Cattuto20071}
  \Name{Cattuto C., Loreto V. \and Pietronero L.}
  \REVIEW{Proc. Natl. Acad. Sci. USA}{104}{2007}{1461}.

\bibitem{ZhangZK201002}
  \Name{Zhang Z.-K. \and Liu C.}
  arXiv: 1003.1931.

\bibitem{Szomszor2007}
  \Name{Szomszor M., Cattuto C., Alani H., OHara K., Baldassarri A., Loreto V. \and Servedio V. D. P.}
  \emph{Proc. the 4th Euro. Semantic Web Conf.} (Innsbruck, Austria) 2007, pp. 71-84.

\bibitem{Schenkel2008}
  \Name{Schenkel R., Crecelius T., Kacimi M., Michel S., Neumann T., Parreira J. X. \and Weikum G.}
  \emph{Proc. the 31st Annual Intl. ACM SIGIR Conf. Res. Dev. Info. Retr.} (ACM Press, New York) 2008, pp. 523-530.



\bibitem{ZhangZK201001}
  \Name{Zhang Z.-K., Zhou T. \and Zhang Y.-C.}
  \REVIEW{Physica A}{389}{2010}{179}.

\bibitem{ShangMS2010}
  \Name{Shang M.-S., Zhang Z.-K., Zhou T. \and Zhang Y.-C.}
  \REVIEW{Physica A}{389}{2010}{1259}.

\bibitem{ShangMS2009}
  \Name{Shang M.-S. \and Zhang Z.-K.}
  \REVIEW{Chin. Phys. Lett.}{26}{2009}{118903}.

\bibitem{Schein2001}
  \Name{Schein A.I., Popescul A., Ungar L.H. \and Pennock D.M.}
  \emph{Proc. 2001 SIGIR Workshop Recomm. Syst.} (New Orleans, LA) 2001.

\bibitem{Schein2002}
  \Name{Schein A. I., Popescul A., Ungar L. H. \and Pennock D. M.}
  \emph{Proc. 25th Annual Intl. ACM SIGIR Conf. Research and Development in Information Retrieval} (ACM Press, New York) 2002, pp. 253-260.

\bibitem{ParkST2006}
  \Name{Park S.T., Pennock D., Madani O., Good N. \and DeCoste D.}
  \emph{Proc. 12th ACM SIGKDD Intl. Conf.  Knowledge Discovery and Data Mining} (ACM Press, New York) 2002, pp. 705-711.



\bibitem{ZhouT200802}
  \Name{Zhou T., Su R.-Q., Liu R.-R., Jiang L.-L., Wang B.-H. \and Zhang Y.-C.}
  \REVIEW{New J. Phys.}{11}{2009}{123008}.


\bibitem{Lambiottea2006}
  \Name{Lambiottea R. \and Ausloosb M.}
  \REVIEW{Eur. Phys. J. B}{50}{2006}{183}.

\bibitem{Eagle2010}
  \Name{Eagle N., Macy M. \and Claxton R.}
  \REVIEW{Science}{328}{2010}{1029}.

\bibitem{Shepitsen2008}
  \Name{Shepitsen A., Gemmell J., Mobasher B. \and Burke R.}
  \emph{Proc. the 2008 ACM Conf. Recomm. Syst.} (ACM Press, New York) 2008, pp. 259-266.

\end{thebibliography}
\end{document}